\documentclass[pre,twocolumn]{revtex4}
\usepackage{psfrag}
\usepackage{amssymb,amsmath,amsthm}
\usepackage[dvips]{graphicx}
\usepackage{verbatim}

\begin{document}
\title{Quantifiers for randomness  of chaotic pseudo random number generators}
\author{L. De Micco}\thanks{CONICET}
\email{lucianadm55@hotmail.com}
\author{H. A. Larrondo}\thanks{CONICET}\email{larrondo@fi.mdp.edu.ar}
\affiliation{Departamento de F\'isica, Facultad de Ingenier\'{\i}a,
Universidad Nacional de Mar del Plata, Juan B. Justo 4302, 7600 Mar
del Plata, Argentina.}
\author{A Plastino}\thanks{CONICET}
\email{plastino@fisica.unlp.edu.ar}\affiliation{Instituto de
F\'{\i}sica, Facultad de Ciencias Exactas, Universidad Nacional de
La Plata, C.C. 727,  1900 La Plata, Argentina.}
\author{O. A. Rosso}\thanks{CONICET}
\email{oarosso@fibertel.com.ar} \affiliation{Centre for
Bioinformatics, Biomarker Discovery and Information-Based Medicine,
and Hunter Medical Research Institute. \\
School of Electrical Engineering and Computer Science,
The University of Newcastle. \\
University Drive, Callaghan NSW 2308, Australia.}
\affiliation{Instituto de C\'alculo,
Facultad de Ciencias Exactas y Naturales, Universidad de Buenos
Aires, 1428 Ciudad Universitaria, Buenos Aires, Argentina.}

\begin{abstract}
We deal with randomness-quantifiers and concentrate  on their
ability do discern the hallmark of chaos in time-series used in
connection with  pseudo random number generators (PRNG). Workers
in the field are  motivated to use chaotic maps for generating
PRNGs because of the simplicity of their implementation. Although
there exist very efficient general-purpose benchmarks for testing
PRNGs, we feel that the analysis provided here sheds additional
didactic light on the importance of the main statistical
characteristics of a chaotic map, namely,  i) its invariant
measure and ii) the mixing constant. This is of help in answering
two questions that arise in applications, that is, (1)
\textit{which is the best PRNG among the available ones?} and (2)
If a given  PRNG turns out  not to be good enough and a
randomization procedure must still be applied to it,
\textit{which is the best applicable randomization procedure?}.
Our answer  provides a comparative analysis of several quantifiers
advanced in the extant literature.

\end{abstract}



\maketitle

\section{Introduction}
\label{sec:INTRO}

Chaos theory started more than thirty years ago and changed
 our world view regarding the role of randomness and
determinism. As  the statistical characteristics of chaotic
systems were better understood
\cite{Lasota1994,Setti2002,Beck1997,Brown1996} a wide variety of
situations emerged  in which chaos, instead of stochastic systems,
became a ``controller of noise" .

Chaos illustrates the rather striking fact that complex behavior
arises from simple rules when nonlinearities are present. Since
simple chaotic maps are capable to generate stochastic-like signals,
implementations based on chaotic systems are usually less involved
than those based in more complex algorithms
\cite{Stojanovski2001,Stojanovski2001b,Kocarev2003}. One tryes to
apply this notion to generate PRNGs because random numbers are
widely used not only in cryptography and Monte Carlo applications
but in less obvious applications \cite{Ecuyer1994,Pecora1990,
Kocarev1995,Hidalgo2001,Fernandez2003}. We mention just  a couple of
them: 1) in spread spectrum techniques, a binary signal is mixed
with a random number sequence to spread the spectrum over a wider
frequency range. Using different random number sequences it is
possible to share a communication channel among several users
\cite{Dinan1998, Mazzini1997a, Shan2006,DeMicco2007B}. Reduction of
electromagnetic interference is another important benefit of the
spread spectrum effect \cite{Setti2000,Callegari2002A}; 2) Consider
a low frequency signal immersed in a high frequency digital noise.
Sampling at time intervals defined by a random number sequence, the
resultant signal becomes filtered without using any coil or
capacitor that are expensive, specially in power systems
\cite{Petrocelli2007}.

Truly random numbers are not attainable from computers and it is
likely that we will ever be able to get them from ``natural"
sources, since one commonly assumes that any system is governed by
underlying physical rules and consequently it is deterministic. A
successful strategy to build up a PRNG is to start with the time
series of a simple nonlinear chaotic map and to apply to it an
adequate  \textit{randomizing procedure} so as to
``heighten/boost" its stochastic nature. Such strategy requires a
quantitative evaluation of the improvement achieved after
effecting the procedure. In \cite{Gonzalez2005} the Statistical
Complexity Measure originally proposed by L\'opez Ruiz et al.
\cite{Lopez1995} and later modified by Lamberti et al.
\cite{Lamberti2004} was used to quantify the effectiveness of such
randomizing modus operandi when applied to a Lorenzian
chaotic system. It was also shown there that a widely employed
course of action -the mixing of two chaotic signals- is not
effective in this respect,  contrary to what one might expect. In
this vein it is important to note that general-purpose tests
available in the open literature \cite{tests} are not designed
taking into account the particular characteristics of a chaotic
map. Instead, one can appreciate in \cite{Rosso2007C} the fact
that the deterministic nature of chaotic dynamics leaves special
manifestations in the associated time series that can be revealed
only by recourse to adequate statistical quantifiers.

In \cite{Demicco2008} chaotic maps were randomized by means of two
different randomizing procedures, namely, {\it discretization} and
{\it skipping}. The idea of concocting an ``information" plane,
called the entropy-complexity plane, was advanced in order to use
it as a means to  ascertain the effectiveness of each of these two
modus operandis. The main difference of this information plane with other
Complexity-Entropy diagrams is the joint use in it of two
different probability distribution functions, both associated to
the pertinent time series.

Other important tools at our disposal are to be mentioned as well.
In a recent and excellent report, Marwan {\it et al.\/} reviewed
applications of so-called Recurrence Plots
for a wide variety of fields and endeavors. They also proposed
several measures to quantify the recurrence plots' characteristics
\cite{Marwan2007}. Additionally, two useful information-theoretic
quantifiers of randomness, the \textit{rate entropy} and the
\textit{Excess Entropy} were proposed in \cite{Feldman2008} as
coordinates of a Complexity-Entropy diagram.

In the present work we explore combinations of all the above
mentioned quantifiers with the purpose to answer the following
questions: 1) \textit{among several chaotic maps, just  which is
the one that generates the best time series?}; 2) \textit{which is
the best strategy -\textbf{Discretization} or \textbf{Skipping}-
to randomize a given chaotic time series?}. The ensuing
quantifiers-testing will be made by means of two representative
chaotic maps (and their iterates).

The paper is organized as follows:  the statistical properties of
a chaotic map and its iterates is reviewed in section
\ref{sec:CHAO}. Section \ref{sec:QUANTI} describes each of the
analyzed quantifiers. Section \ref{sec:RESUL} deals with results
for two representative maps and, finally, conclusions are
presented in Section \ref{sec:SUMM}.

\section{Statistical Properties of a Chaotic Map}
\label{sec:CHAO}

Let $f$ be a chaotic map on the interval $[0,1]$. Suppose the map
has an invariant measure $\mu(x)$. Then the map is \textit{ergodic}
if for any integrable test function $Q(x)$, and for an arbitrary
initial condition $x_0$ (up to a set of zero $\mu$-measure), the
time average is equal to the ensemble average:
\begin{equation}\label{eq:ERGO}
    \overline{Q}=\langle Q\rangle\;.
\end{equation}
Equation (\ref{eq:ERGO}) is a consequence of the famous Birkhoff
ergodic theorem \cite{Cornfeld1982}. \textit{Mixing} is an even
stronger requirement  than ergodicity. A map is called
``\textit{mixing}" if any smooth initial probability density
$\rho(x)$ converges to the invariant measure $\mu(x)$ after enough
successive iterations. Mixing implies ergodicity. The reverse,
however, is not true \cite{Beck1990}.

There exists an equivalent definition of mixing via
\textit{correlation functions}. Let $\phi_1(x)$ and $\phi_2(x)$ be
two integrable test-functions and define the generalized
correlation function of the map $f$ by
\begin{equation}\label{eq:CORRE}
    C(\phi_1,\phi_2,n)=\lim_{J\rightarrow\infty}\frac{1}
    {J}\sum_{j=0}^{J-1}\phi_1(x_{j+n})\phi_2(x_{j})-\langle\phi_1\rangle \langle
    \phi_2\rangle\;,
\end{equation}
where
\begin{equation}\label{eq:MEAN}
    \langle\phi_i\rangle= \lim_{J\rightarrow\infty}\frac{1}
    {J}\sum_{j=0}^{J-1}\phi_i(x_{j})\;.
\end{equation}
The map is mixing if, for arbitrary $\phi_1$ and $\phi_2$,
\begin{equation}\label{eq:MIX}
   \lim_{n\rightarrow\infty}C(\phi_1,\phi_2,n)=0\;.
\end{equation}

Let us stress that it is not easy to prove that $f$ is a
\textit{mixing map} because the mixing condition given in Eq.
(\ref{eq:MIX}) must be fulfilled for \textit{arbitrary} test
functions. Formally,  every mixing map $f$ has an associated
\textit{Perron-Frobenius operator} $\mathcal{L}$ \cite{Beck1990}
that determines the time evolution of any initial density
$\rho_0(x)$ towards the invariant measure $\mu(x)$:
\begin{equation}\label{eq:PERRON1}
   \rho_{n+1}=\mathcal{L}[\rho_n] \;.
\end{equation}
The explicit formal expression for the Perron-Frobenius operator for
a one-dimensional map $f$ is given by \cite{Beck1990}
\begin{equation}\label{eq:PERRON2}
   \mathcal{L}[\rho_{y}]=\sum_{x\epsilon f^{-1}(y)}
   \frac{[\rho_0(x)]}{|f'(x)|}\;.
\end{equation}
This operator $\mathcal{L}$ has a set of eigenfunctions
$\psi_\alpha(x)$ and eigenvalues $\eta_\alpha$. The invariant
measure $\mu(x)$ is the eigenfunction corresponding to the largest
eigenvalue $\eta_0=1$. The full set of eigenfunctions and
eigenvalues may be used as a basis to express any density:
\begin{eqnarray}
\nonumber   \rho_{0}(x) &=& \sum_\alpha c_\alpha \psi_\alpha(x) \ , \\
 \rho_{n}(x) &=& \mathcal{L}^{n}\rho_{0}(x) = \sum_\alpha \eta_\alpha^{n} c_\alpha = c_0
 \psi_0(x)+R_n \ .
\end{eqnarray}
The eigenvalue with the second largest absolute value, $\eta_1$, has
a "distinguished" physical meaning: it is related with the
\textit{mixing constant} $r_{mix}$ that governs the relaxation of
\textit{exponentially mixing} maps:
\begin{equation}\label{eq:RMIX}
    |R_n|\sim |\eta_1|^{n}\sim \exp(-n/r_{mix})\;.
\end{equation}

The invariant measure $\mu(x)$ gives the histogram of the time
series and the ideal PRNG must have $\mu(x)=const$. The mixing
constant $r_{mix}$ gives the transient characteristic time
\cite{DeMicco2007B,Petrocelli2007,Mazzini1997a,Rovatti2004a,Rovatti2004b}
and its ideal value is $r_{mix}=0$. In many applications of PRNGs
both the invariant measure and the mixing constant are relevant.

The analytical expression of the invariant measure $\mu(x)$ of a
given map $f$ is usually not known. Exceptions are the logistic map
in full chaos, and the piecewise-linear maps. The mixing constant
$r_{mix}$ has been analytically obtained only for piecewise-linear
maps. For other maps it must be numerically obtained by means of a
piecewise linear approximation of the map
\cite{Lasota1994,Beck1997}.

It is then obviously convenient to have quantifiers for measuring
the uniformity of the invariant measure $\mu(x)$, and the mixing
constant $r_{mix}$, for any chaotic map. These quantifiers are
useful to compare time series coming from different chaotic maps
and also to assess the improvements produced by
\textit{randomization procedures}.

It is possible to show that the invariant measure of $f^{d}$ is
identical to the invariant measure of $f$. Also, the mixing
constant $r_{mix}$ for $f^{d}$ is lower that the mixing constant
for $f$. The iteration of a map is one of the randomization
procedures proposed in the literature, being used to diminish
$r_{mix}$. This procedure is also known as \textbf{Skipping}
because iterating is tantamount to skipping values in the original
time series, which does not change $\mu(x)$ and, consequently, is
not conductive to a randomization of  chaotic maps with
$\mu(x)\neq const$. In this paper we will use ``skipping" as a
method of quantifier-analysis.
\section{Quantifiers for the invariant measure and mixing constant}
\label{sec:QUANTI}
In this section we review several quantifiers proposed for
measuring the main statistical properties of chaotic PRNGs. The
quantifiers are classified according to their origin into three
classes: (1) quantifiers based on Information Theory
\cite{Lopez1995,Rosso2007C,Lamberti2004}; (2) quantifiers based on
Recurrence Plots \cite{Eckmann1987,Marwan2007}; (3) quantifiers
based on intrinsic computation \cite{Feldman2008}.

\subsection{Quantifiers based on Information Theory}

They are appropriate functionals of the probability distribution
function (PDF). Let $\{x_i\}$ be the time series under analysis,
with length $M$. There are infinite possibilities to assign a PDF
to a given time series, a subject that will be given due
consideration below. In the meantime, suppose  that the PDF is
discrete and  is given by $P=\{p_i; i=\cdots, N\}$. One  defines
 then various quantities, namely, %
\begin{enumerate}
  \item Normalized Shannon Entropy $H[P]$.
   Let $S[P]$ be the Shannon Entropy
  \begin{equation}
  \label{eq:ssha}
   S[P]=-\sum_{i=1}^{N}p_i\;\ln({p_i}) \ .
  \end{equation}
  Is is well known that the maximum $S_{max}=\ln(N)$ is obtained for
$P_e=\{1/N, \cdots,  1/N\}$, that is,
  the uniform PDF. A ``normalized" entropy $H[P]$ can also be defined in the fashion
  \begin{equation}
  \label{eq:hsha}
   H[P] = S[P] / S_{max} \ .
  \end{equation}
  \item Statistical Complexity Measure. A full discussion about Statistical Complexity Measures exceeds
   the scope of this presentation. For a comparison amongst different complexity measures see
   the excellent paper by Wackerbauer et al. \cite{Wackerbauer1994}.
   In this paper we adopt the definition by L\'opez Ruiz-Mancini-Calbet seminal paper \cite{Lopez1995}
   with the modifications advanced in \cite{Lamberti2004} so as to ensure that the concomitant
   SCM-version becomes
   \textit{(i)} able to grasp essential details of the dynamics,
   \textit{(ii)} an intensive quantity and,
   \textit{(iii)} capable of discerning both among different degrees of periodicity and
   chaos \cite{Rosso2007C}.
   The ensuing measure, to be referred to as the intensive statistical complexity, is a functional $C[P]$
   that reads
   \begin{equation}
   \label{eq:inten}
   C[{P}]=Q_{J}[{P,P_e}]\cdot H[{P}]  ,
   \end{equation}
   where $Q_{J}$ is the ``disequilibrium", defined in terms of the so-called extensive Jensen-Shannon divergence
    (which induces a squared metric) \cite{Lamberti2004}. One has
   \begin{equation}
   \label{eq:disequi}
   Q_{J}[{P,P_e}]= Q_0  \cdot \{S[(P+P_e)/2]-S[P]/2-S[P_e]/2 \},
   \end{equation}
   with $Q_0$ a normalization constant ($0 \le Q_{J} \le 1$) that reads
   \begin{equation}
   \label{eq:q0j}
    Q_0=-2 \left\{ \left( \frac{N+1}{N} \right) \ln(N+1) - 2 \ln(2N) + \ln N \right\}^{-1} .
   \end{equation}
   We see that the disequilibrium $Q_J$ is an intensive quantity that
   reflects on the systems's ``architecture", being different from zero
   only if there exist ``privileged", or ``more likely" states among
   the accessible ones. $C[P]$ quantifies the presence of
   correlational structures as well \cite{Martin2003,Lamberti2004}. The
   opposite extremes of perfect order and maximal randomness possess no
   structure to speak of and, as a consequence, $C[P]=0$. In
   between these two special instances a wide range of possible degrees
   of physical structure exist, degrees that should be reflected in the
   features of the underlying probability distribution.
   In the case of a PRNG the ``ideal" values are $H[{P}]=1$ and $C[{P}]=0$.
\end{enumerate}
As pointed out above, $P$ itself is not a uniquely defined object
and several approaches have been employed in the literature so as to
``extract" $P$ from the given time series. Just to mention some
frequently used extraction procedures: {\it a)\/} time series
histogram \cite{Martin2004}, {\it b)\/} binary symbolic-dynamics
\cite{Mischaikow1999}, {\it c)\/} Fourier analysis
\cite{Powell1979}, {\it d)\/} wavelet transform
\cite{Blanco1998,Rosso2001}, {\it e)\/} partition entropies
\cite{Ebeling2001}, {\it f)\/} permutation entropy
\cite{Bandt2002a,Keller2005}, {\it g)\/} discrete entropies
\cite{Amigo2007B}, etc. There is ample liberty to choose among them.
In \cite{Demicco2008} two probability distribution were proposed as
relevant for testing the uniformity of  $\mu(x)$ and the mixing
constant: (a) a $P$ based on time series' histograms and (b) a $P$
based on ordinal patterns (permutation ordering) that derives from
using the Bandt-Pompe method \cite{Bandt2002a}.

For extracting $P$ via the histogram divide the interval $[0,1]$
into a finite number $N_{bin}$ of non overlapping subintervals
$A_i$: $[0,1]=\bigcup_{i=1}^{N_{bin}} A_i$ and $A_i\bigcap
A_j=\emptyset~\forall i\neq j$. Note that $N$ in eq. (\ref{eq:ssha})
is equal to $N_{bin}$. Of course, in this approach the temporal
order of the time-series plays no role at all. The quantifiers
obtained via the ensuing PDF are called in this paper $H^{(hist)}$
and $C^{(hist)}$. Let us stress that for time series within a finite
alphabet it is relevant to consider an optimal value of $N_{bin}$
(see i.e. \cite{Demicco2008}).

In extracting $P$ by recourse to the Bandt-Pompe method  the
resulting probability distribution $P$ is based on the details of
the attractor-reconstruction procedure. {\it Causal information\/}
is, consequently,  duly incorporated into the construction-process
that yields $P$. The quantifiers obtained via the ensuing PDF are
called in this paper $H^{(BP)}$ and $C^{(BP)}$. A notable
Bandt-Pompe result consists in getting a clear improvement in the
quality of  information theory-based quantifiers
\cite{Larrondo2005,Larrondo2006,Kowalski2007,Rosso2007C,Rosso2007A,Rosso2007B,Zunino2007A,Zunino2007B}.

The extracting procedure is as follows. For the time-series $\{x_t :
t=1,\cdots,M \}$ and an embedding dimension $D > 1$, one looks for
``ordinal patterns" of order $D$
\cite{Bandt2002a,Keller2003,Keller2005} generated by
\begin{equation}
\label{eq:vectores}
(s)\mapsto \left(x_{s-(D-1)},x_{s-(D-2)},\cdots,x_{s-1},x_{s}\right) \ ,
\end{equation}
which assign to each ``time $s$" a $D$-dimensional vector of values
pertaining to the times $s, s-1,\cdots,s-(D-1)$. Clearly, the
greater the $D-$value, the more information on ``the past" is
incorporated into these vectors. By the ``ordinal pattern" related
to the time $(s)$ we mean the permutation $\pi=(r_0, r_1, \cdots, r_{D-1})$ of $(0, 1, \cdots, D-1)$
defined by
\begin{equation}
\label{eq:permuta}
x_{s-r_{D-1}}\le x_{s-r_{D-2}}\le\cdots\le x_{s-r_{1}}\le x_{s-r_0}.
\end{equation}
In order to get a unique result we consider that  $r_i <r_{i-1}$ if
$x_{s-r_{i}}=x_{s-r_{i-1}}$. Thus, for all the $D!$ possible
permutations $\pi$ of order $D$, the probability distribution
$P=\{p(\pi)\}$ is defined by
\begin{equation}
\label{eq:frequ}
p(\pi)~=~ \frac{\sharp \{s|s\leq M-D+1; (s) \quad \texttt{has type}~\pi\}}{M-D+1}.
\end{equation}
In the last expression the symbol $\sharp$ stands for ``number".

The advantages of the Bandt-Pompe method reside in {\it a)\/} its
simplicity, {\it b)\/} the associated extremely fast
calculation-process, {\it c)\/} its robustness in presence of
observational and dynamical noise, and {\it d)\/} its invariance
with respect to nonlinear monotonous transformations. The
Bandt-Pompe's methodology is not restricted to time series
representative of low dimensional dynamical systems but can be
applied to any type of time series (regular, chaotic, noisy, or
reality based), with a very weak stationary assumption (for $k =
D$, the probability for $x_t < x_{t+k}$ should not depend on $t$
\cite{Bandt2002a}). One also assumes that enough data are
available for a correct attractor-reconstruction. Of course, the
embedding dimension $D$ plays an important role in the evaluation
of the appropriate probability distribution because $D$ determines
the number of accessible states $D!$. Also, it conditions the
minimum acceptable length $M \gg D!$ of the time series that one
needs in order to work with a reliable statistics. In relation to
this last point Bandt and Pompe suggest, for practical purposes,
to work with $3\leq D \leq 7$ with a time lag $\tau = 1$. This is
what we do here (in the present work $D=6$ is used).

\subsection{Quantifiers based on Recurrence Plots:}

Recurrence Plots were introduced by Eckmann {\it et al.\/}
\cite{Eckmann1987} so as to to visualize the recurrence of states
during phase space-evolution. The recurrence plot is a two
dimensional representation in which both axes are time-ones. The
recurrence of a state appearing at two given times $t_i,\,t_j$ is
pictured in the two-dimensional graph by means of either black or
white dots, where a black dot signals a recurrence. Of course only
periodic continuous systems will have exact recurrences. In any
other case one detects only approximate recurrences, up to an
error $\epsilon$. The so-called recurrence function can be
mathematically expressed as:
\begin{equation}
\label{eq:rp}
\mathbf{R}(i,j)=\Theta\left(\varepsilon-\|\overrightarrow{x}(i)-
\overrightarrow{x}(j)\|\right),\quad
\end{equation}
with $\overrightarrow{x}(i)\in {\Re}^{m}$ and $i,j=1,\cdots,N$.
Being $N$ the number of discrete states $\overrightarrow{x}(i)$
considered, $\|\bullet\|$ is a norm, and $\Theta(\bullet)$ is the
Heaviside step function.

In the particular case of the PRNGs analyzed in this paper only
$1{\mathcal D}$ series are considered but the recurrence
function-idea can be extended to ${\mathcal D}$-dimensional phase
spaces or even to suitably reconstructed embedding phase spaces.
Of course, the visual impact produced by the recurrence plot is
 insufficient to compare the quality of different PRNGs, because of
the ``small scale" structures that may be present in our scenario.
Several kind of measures have been defined to quantify these small
scale structures \cite{Marwan2007}, each measure being a
functional of the recurrence function (Eq. (\ref{eq:rp})). In this
paper two kind of Recurrence Plot-measures are considered, namely,
\begin{enumerate}
   \item Measures based on the recurrence density (measured by  the number of points in the Recurrence Plot).
    In this paper we use the \textit{Recurrence Rate} ($RR$), given by:
    \begin{equation}
    \label{eq:RR}
    RR(\varepsilon)=\frac{1}{N^{2}}\sum_{i,j=1}^{N}\textbf{R}_{ij}(\varepsilon)\;.
    \end{equation}
    Note that in the limit $N\rightarrow\infty$,  $RR$ is the probability
    that a state recurs to its $\varepsilon$-neighborhood in phase space.
    For PRNGs the ideal value would be $RR=0$. But in practice, if no points are
    to be found in the Recurrence Plot, a larger $\varepsilon$ must be
    adopted  in order that the quantifier may make sense.
    \item Diagonal Measures: these are measures related to the histogram $P(\varepsilon,l)$ of diagonal line lengths, given by
    \begin{eqnarray}
    \label{eq:PDIAG}
    P(\varepsilon,l)=\sum_{i,j=1}^{N}\left[1-\textbf{R}_{i-1,j-1}(\varepsilon)\right]\left[1-\textbf{R}_{i+l,j+l}(\varepsilon)\right]  \cdot \nonumber\\
    \prod_{k=0}^{l-1}\textbf{R}_{i+k,j+k}(\varepsilon)\;.
    \end{eqnarray}
    Processes with uncorrelated or weakly correlated behavior originate no
    (or just  very short) diagonals, whereas deterministic processes give
    rise to ``long" diagonals and smaller amount of single, isolated recurrence points. In
    this paper three measures based on the statistics of diagonal lines are considered:
    \begin{enumerate}
        \item  the \textit{deterministic quantifier} DET, the ratio of
         recurrence points that form diagonal structures of at least length
         $l_{min}$ to all recurrence points
         \begin{equation}
         \label{eq:DET}
         DET=\frac{\sum_{l=l_{min}}^{N} l \cdot P(\varepsilon,l)}
         {\sum_{l=1}^{N}  l \cdot P(\varepsilon,l)}\;;
         \end{equation}
         \item The\textit{ average diagonal line length} L
         \begin{equation}
         \label{eq:L}
         L=\frac{\sum_{l=l_{min}}^{N}  l \cdot P(\varepsilon,l)}
         {\sum_{l=l_{min}}^{N}P(\varepsilon,l)}\;;
         \end{equation}
         \item The \textit{entropy} $ENTR$ given by
         \begin{equation}
         \label{eq:ENTR}
         ENTR=-\sum_{l=l_{min}}^{N}P(\varepsilon,l) \ln P(\varepsilon,l)\;.
         \end{equation}
    \end{enumerate}
\end{enumerate}

\subsection{Quantifiers based on intrinsic computation}
We consider in this paper two quantifiers introduced in
\cite{Feldman2008}, i.e.,  the \textit{entropy rate} $h_{\mu}$ and
the \textit{entropy excess} $\textbf{E}$. They are defined for
time series with a finite alphabet $\mathcal{A}$, which is  not a
limitation because the $x_i$'s may be thought of as real numbers
only in analytical studies. In any practical case they are in fact
\textit{floating point numbers} and, consequently, they have only
a finite number of allowed $A-$values. A subsequence $s^{L}=\{x_i,
x_{i+1}, \cdots ,x_{i+L}\}$ is called an $L$-block. Let $P(s^{L})$
denotes the probability of a particular $L$-block. Then the block
entropy $H(L)$ is:
\begin{equation}
\label{eq:hblock}
H(L)\equiv -\sum_{s^{L}}\; P(s^{L})\;log_2\,P(s^{L}) \ .
\end{equation}
The sum runs over all possible blocks of length $L>0$ and
$H(0)\equiv 0$ by definition. For stationary processes and
sufficiently large $L$, $H(L)\sim L$. On the other hand, the
\textit{entropy rate} $h_{\mu}$ is defined as
\begin{eqnarray}\label{eq:rateEntropy}
  \nonumber h_{\mu}(L) &=& \frac{H(L)}{L} \ , \\
  h_{\mu}&=&\lim_{L\rightarrow\infty}\quad  h_{\mu}(L) \ .
\end{eqnarray}
The \textit{entropy rate} is also known as the \textit{metric
entropy} in dynamical systems' theory and it is equivalent to the
\textit{thermodynamic entropy density} familiar from equilibrium
statistical mechanics \cite{Feldman2008}. The entropy rate
provides a reliable and well understood measure of the randomness
or disorder intrinsic to a process. However, this tell us little
about the underlying system's organization, structure or
correlations. A measure of system's organization may be obtained
by looking at the manner in which $h_{\mu}(L)$ converges to its
asymptotic value $h_{\mu}$. When only observations over length
$L$-blocks  are taken into account, a process appears to have an
entropy rate of $h_{\mu}(L)$  larger than the asymptotic value of
 $h_{\mu}$. As a result, the process seems to be of a more random nature than it really is by the ``excess" amount of
$h_{\mu}(L)-h_{\mu}$ bits. Summing these entropy over-estimates
over $L$ one obtains the \textit{excess entropy}
\cite{Crutchfield1983}:
\begin{equation}
\label{eq:excessEntropy}
\textbf{E}\equiv\sum_{L=1}^{\infty}\;\left[h_{\mu}(L)-h_{\mu}\right].
\end{equation}

\subsection{Expected behavior for PRNG}
Summing up, the quantifiers to be compared here are: $H^{(hist)}$,
$C^{(hist)}$, $H^{(BP)}$, $C^{(BP)}$, $RR$, $DET$, $ENTR$, $L$,
$h_{\mu}$, and $\textbf{E}$. These quantities should tell us how
good our PRNG is as compared to  the ideal condition
$\mu(x)=const$, $r_{mix}=0$.  $H^{(hist)}$ is the natural
quantifier to measure a non constant $\mu(x)$, with value $1$ for
the ideal PRNG. It does not depend on the order of appearance of a
given time-series event,  but only in the number of times such
event takes place. As for $H^{(hist)}$, it is not able to uncover
any change in $r_{mix}$-values. Thus, to get a good representation
plane we ought to demand  a quantifier that changes with $r_{mix}$
and not with $\mu(x)$. To look for such a kind of  quantifier we
must study $H^{(hist)}$, $C^{(hist)}$, $H^{(BP)}$, $C^{(BP)}$,
$RR$, $DET$, $ENTR$, $L$, $h_{\mu}$, and $\textbf{E}$ as functions
of $r_{mix}$. A family of iterated maps $f^{d}$ may be used to
that end  because they share the same invariant measure and
$r_{mix}$ is a decreasing function of $d$. The best quantifier for
$r_{mix}$ would be that which has maximal variation over the
entire family of maps.

\section{Application to Logistic Map and Three Way Bernoulli Map}
\label{sec:RESUL}
In this Section we present results  for the families of iterates
of two chaotic well-known maps,  the Logistic Map (LOG) and the
Three Way Bernoulli Map (TWB), that  have been selected, among
other possibilities, because they are representative of two
different classes of systems.
\begin{itemize}
\item  LOG is given by:
\begin{equation}
\label{eq:logimap}
    x_{n+1}~=~4~x_n~(1-x_n),
\end{equation}
and its natural invariant density can be exactly determined, being
expressed in the fashion
\begin{equation}\label{eq:logNID}
    \rho_{inv}(x)~=~\frac{1}{\pi \sqrt{x(1-x)}}\;  .
\end{equation}
LOG is paradigmatic because it is representative not only of maps
with a quadratic maximum, but also emerges when the Lorenz
procedure is applied to many continuous attractors with basins
that may be approached with the Lorenz method via a $1{\mathcal
D}$-map (like the Lorenz, Rossler, Colpits ones among others). A
{\it non uniform\/} natural invariant density is an important
feature in this instance \cite{Beck1997}. The ensuing
$r_{mix}$-values are also displayed in Table \ref{tab:tablermix}.
They  have been obtained by recourse to the Transfer Operator
Method, as described in \cite{Beck1997}.
\item TWB is given by
\begin{equation}
x_{n+1}~=~ \left\{ \begin{array}{ll}
3{x_n} & \textrm{if $0\leq x_n\leq 1/3$}\\
3{x_n} -1& \textrm{if $1/3<x_n\leq 2/3$} \\
3{x_n} -2& \textrm{if $2/3<x_n\leq 1$}
\end{array} \right.  \ .
\end{equation}
TWB is representative of the class of piecewise-linear maps as,
for example, the Four Way Tailed Shift Map, the Skew Tent Map, the
Three Way Tailed Shift Map, etc. All these maps share a {\it
uniform\/}
natural invariant density \cite{Beck1997}. The mixing constant 
of the whole family of maps $f^{d}$ is given by
$r^{d}_{mix}=(1/3)^{d}$ (see table \ref{tab:tablermix}).
\end{itemize}

\begin{table}
\begin{tabular}{ccc}
\hline\hline
  $d$     &TWB        &LOG \\
\hline\hline
  1     &0.56789     &0.333333333 \\
  2     &0.31848     &0.111111111 \\
  3     &0.13290     &0.037037037 \\
  4     &0.05788     &0.012345679 \\
  5     &0.03646     &0.004115226 \\
  6     &0.01791     &0.001371742 \\
  7     &0.01152     &0.000457247 \\
  8     &0.00515     &0.000152416 \\
  \hline\hline
\end{tabular}
\caption{$r_{mix}$ as a function of the iteration-order $d$ for LOG
and TWB.}\label{tab:tablermix}
\end{table}

For the evaluation of the different quantifiers  we used files with
$M = 50 \cdot 10^6$ floating point numbers. In the Band-Pompe
approach we consider $D=6$ while for histograms we have taken
$N_{bin} = 2^{16}$. All Recurrence Plots-measures depend on several
parameters:
\begin{itemize}
    \item the dimension $D_e$ of the embedding space. In this paper $D_e=1$.
    \item $\varepsilon$, a parameter crucial so as to  define just when
    recurrences occur. We adopted $\varepsilon=1/(2^{16}-1)$ corresponding to $16$-bits numbers.
    \item $l_{min}$ is the minimum length accepted for diagonal
    lines. $l_{min}=2$ is used in this paper for all diagonal measures except for $L$ ($l_{min}=1$ is used for $L$).
    \item $N$ is the number of values used for each realization.
    In this paper values of $RR$, $DET$, $ENTR$ and $L$ are mean values over $10$ surrogate-series with $N=10000$ data
    each.
\end{itemize}

\begin{figure*}[ptb]
\centering
\begin{minipage}[c]{\textwidth}
\includegraphics[width=0.49\textwidth]{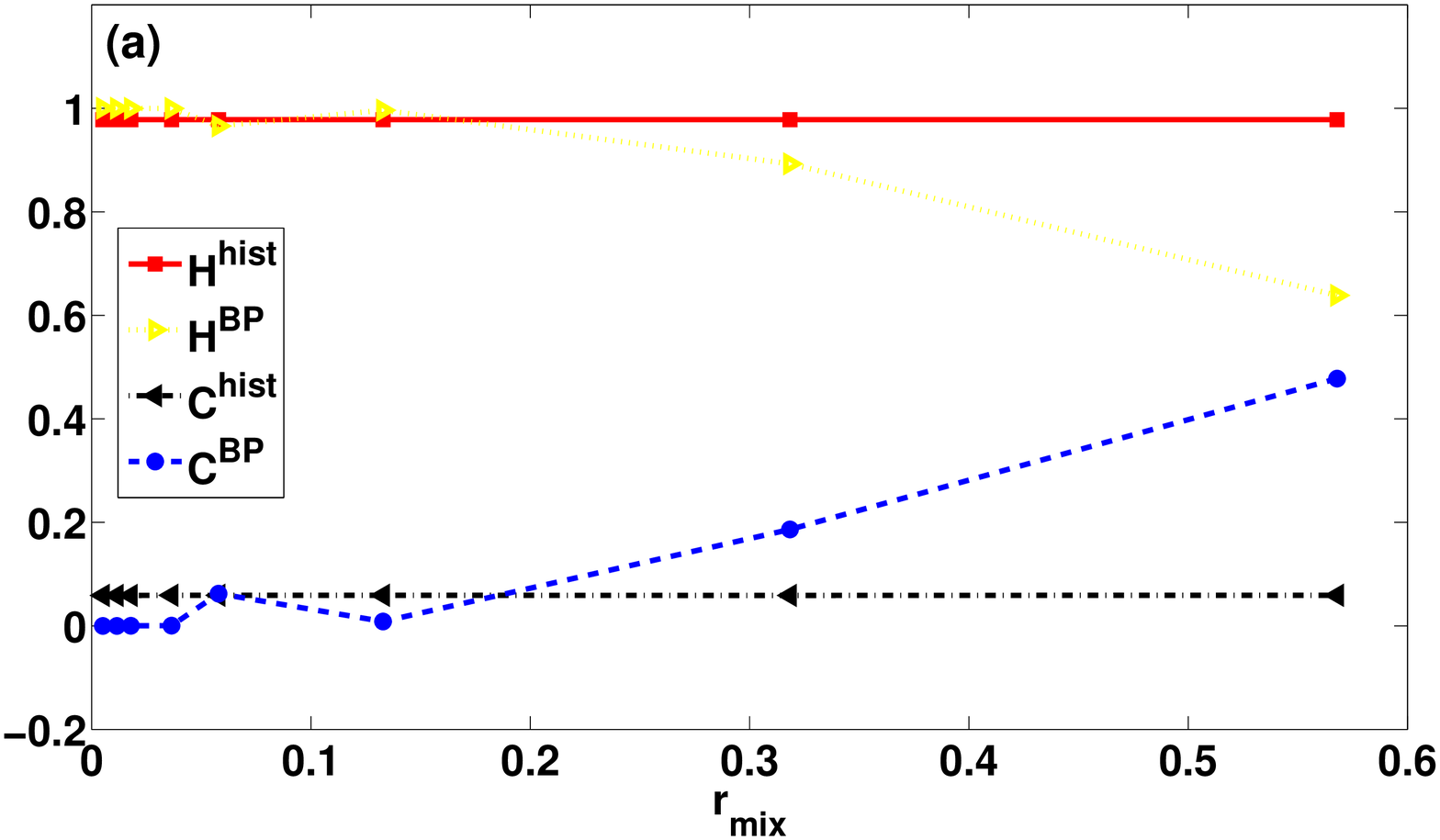}
\includegraphics[width=0.49\textwidth]{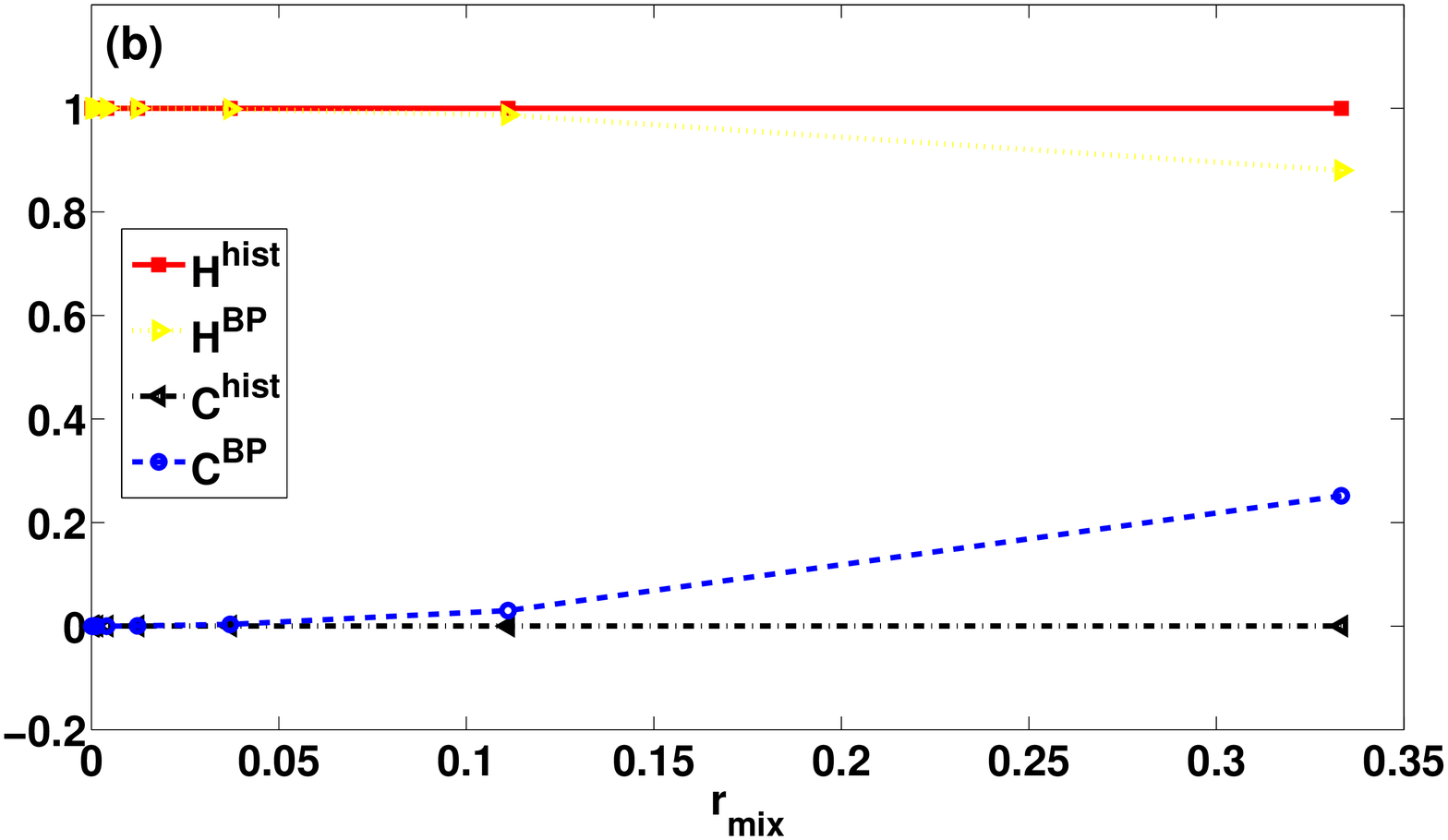}
\caption{Information Theory quantifiers as functions of $r_{mix}$
for: {\it (a)\/} LOG map, {\it (b)\/} TWB map. }
\label{fig:infoQuanti}
\end{minipage}
\end{figure*}

\begin{figure*}[ptb]
\centering
\begin{minipage}[c]{\textwidth}
\includegraphics[width=0.49\textwidth]{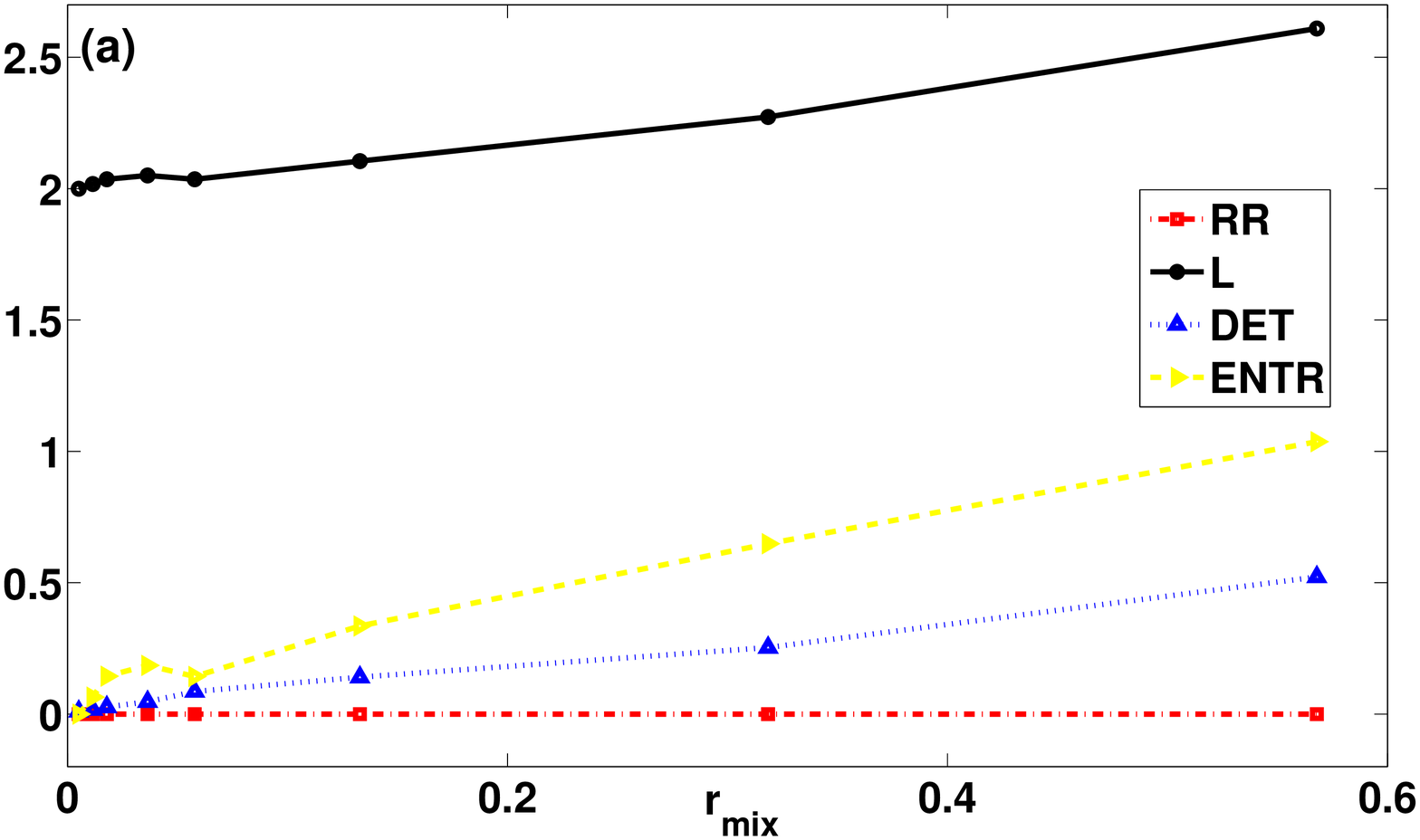}
\includegraphics[width=0.49\textwidth]{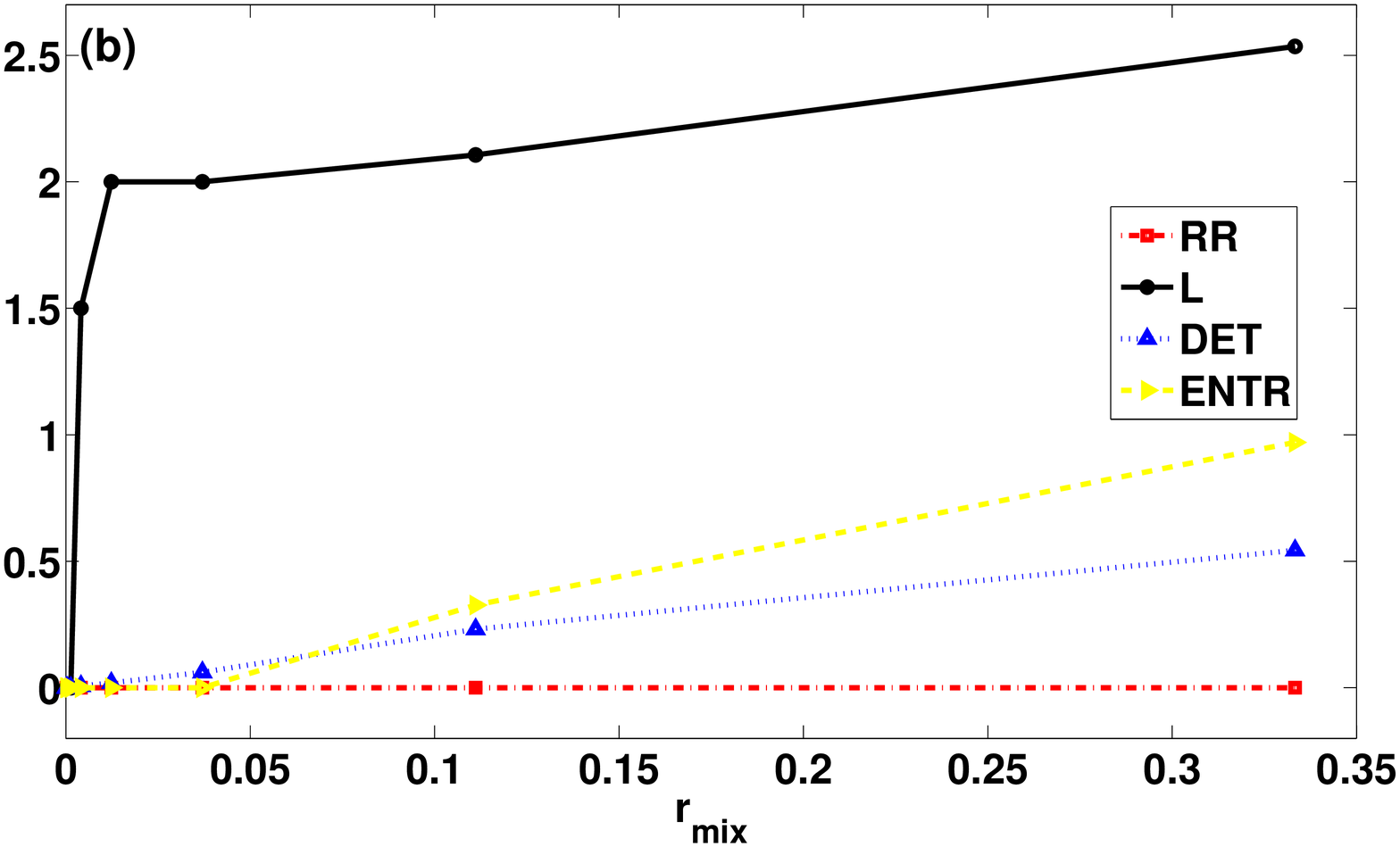}
\caption{Recurrence Plots quantifiers as functions of $r_{mix}$ for:
{\it (a)\/} LOG map, {\it (b)\/} TWB map. } \label{fig:crpQuanti}
\end{minipage}
\end{figure*}

\begin{figure*}[ptb]
\centering
\begin{minipage}[c]{\textwidth}
\includegraphics[width=0.49\textwidth]{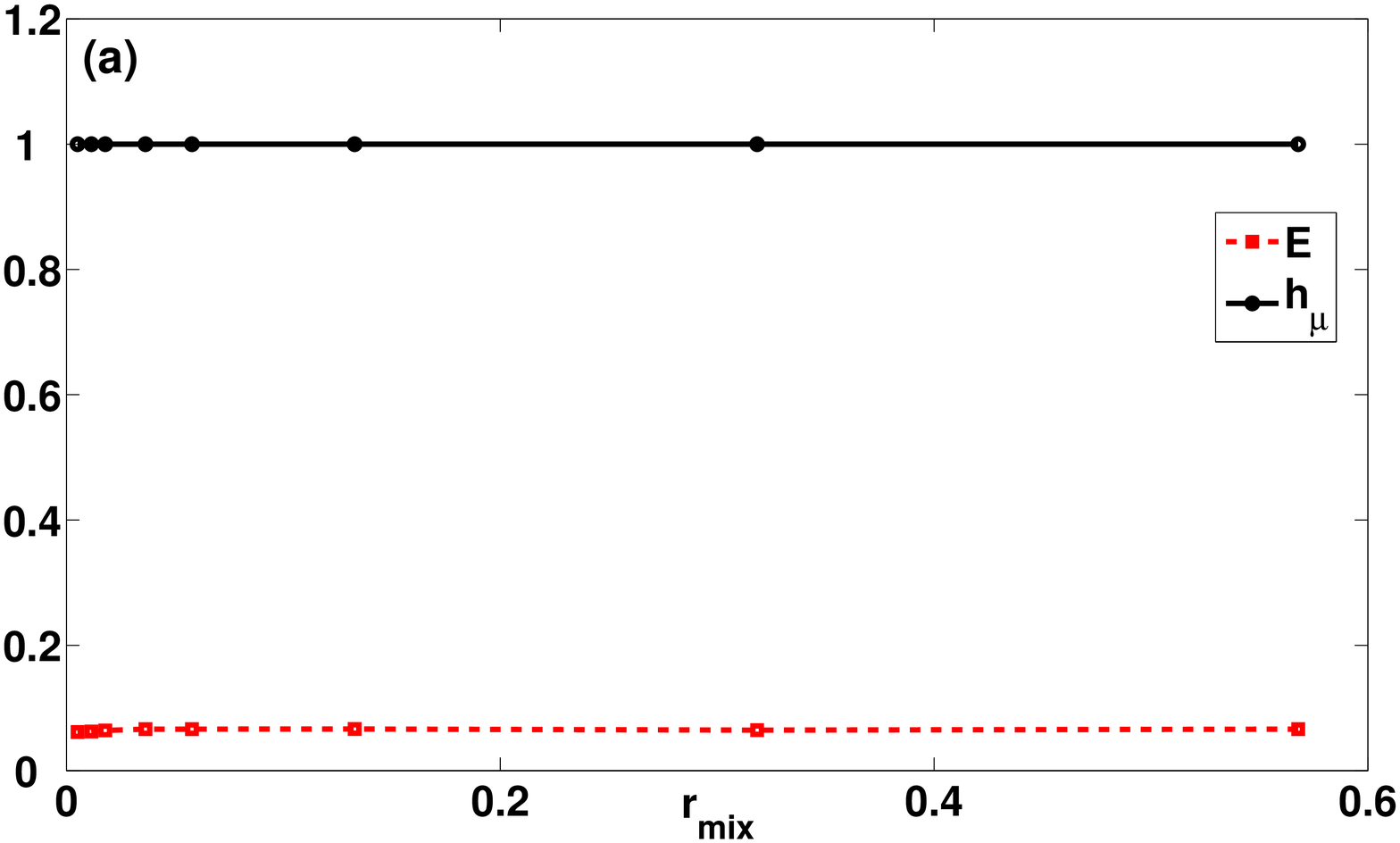}
\includegraphics[width=0.49\textwidth]{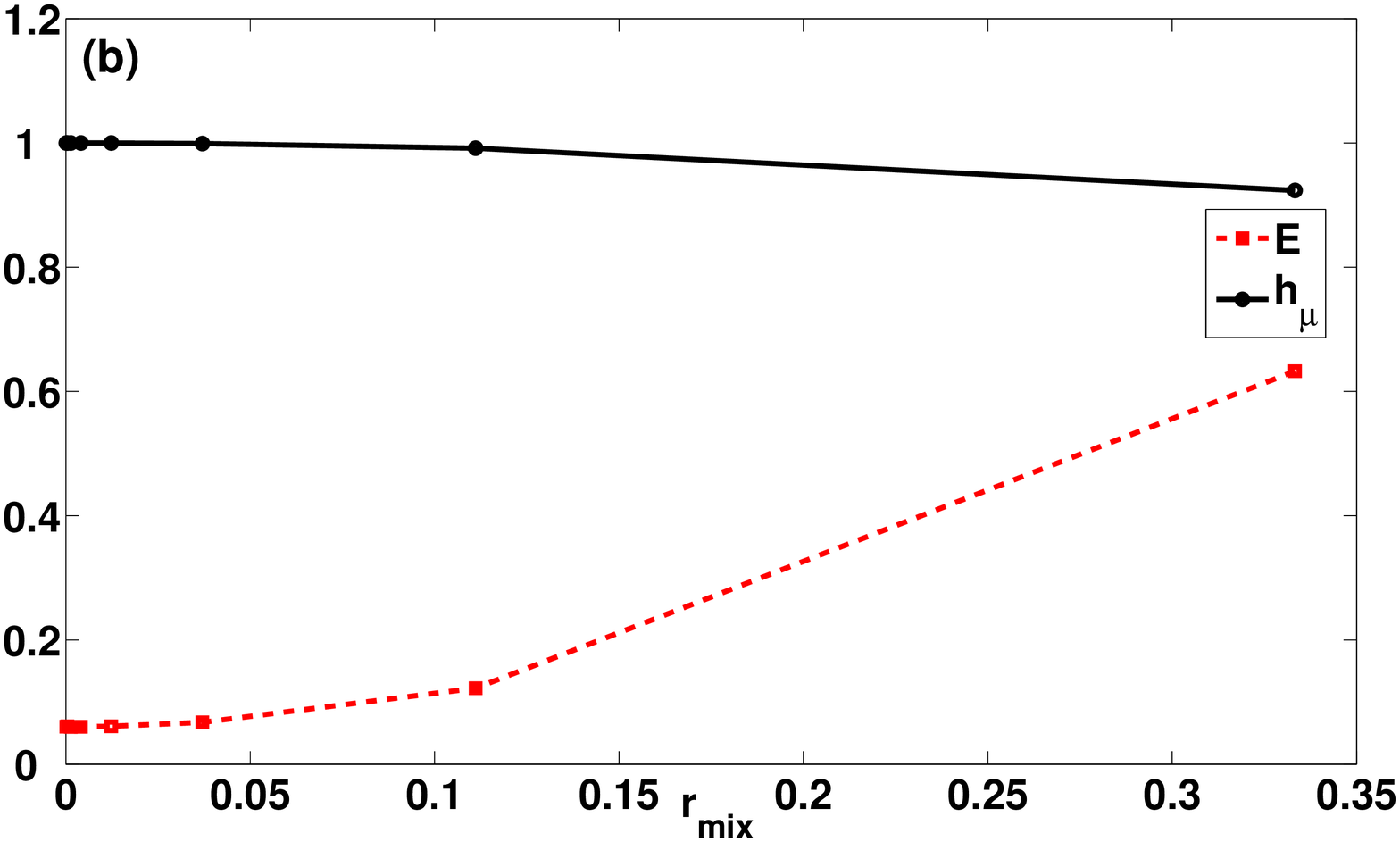}
\caption{Intrinsic Computation quantifiers as functions of $r_{mix}$
for: {\it (a)\/} LOG map, {\it (b)\/} TWB map. }
\label{fig:intriQuanti}
\end{minipage}
\end{figure*}

Figures \ref{fig:infoQuanti}, \ref{fig:crpQuanti}, and
\ref{fig:intriQuanti} illustrate the behavior of all the
quantifiers for the iterates of LOG (Figs. (a)) and the iterates
of TWB (Figs. (b)),  respectively. These figures show that the
following quantifiers are the ones usable for measuring $r_{mix}$,
namely,  $C^{(BP)}$, $DET$, $ENTR$, and $L$. On the other hand,
the following quantifiers depend on the invariant density but they
do not depend on $r_{mix}$: $H^{(hist)}$, $C^{(hist)}$, and $RR$.

Intrinsic computation quantifiers display a completely different
behavior for LOG and for TWB. In LOG both quantifiers have no
dependence with $r_{mix}$, but in TWB, $h_{\mu}$ decreases as
$r_{mix}$ increases while $\mathbf E$ is an increasing function of
$r_{mix}$. Thus, these parameters do not seem to be  convenient
ones.

Comparing LOG with TWB by recourse to  these parameters
shows that TWB is slightly better than LOG. The problem with TWB
and with other piecewise linear maps is they are not realistic
enough and that their implementation is more involved that for
other maps like LOG.

\begin{figure*}[ptb]
\centering
\begin{minipage}[c]{\textwidth}
\includegraphics[width=0.49\textwidth]{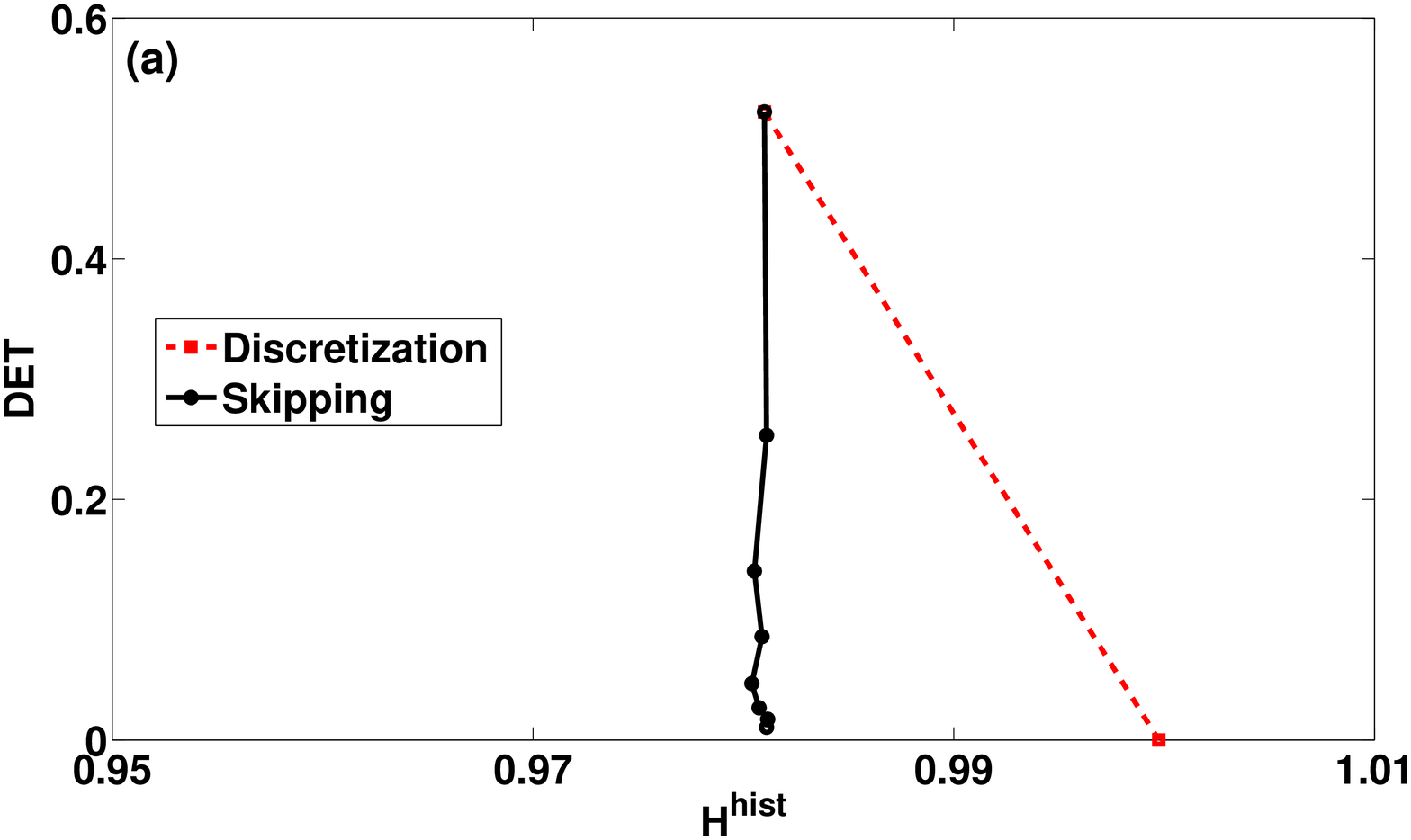}
\includegraphics[width=0.49\textwidth]{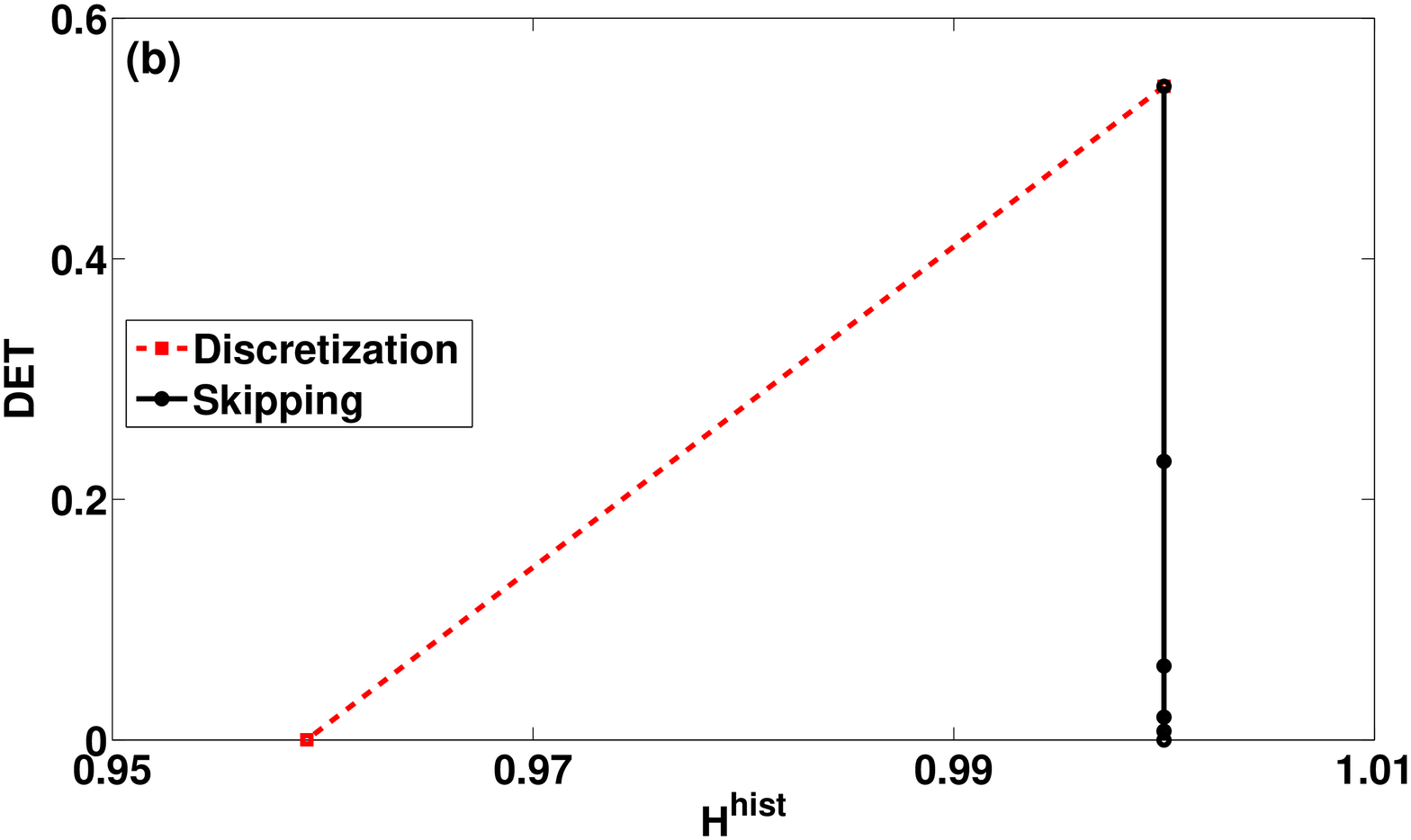}
\caption{$DET$ as a function of $H^{hist}$, as evaluated in
\cite{Demicco2008}, for both randomization procedures applied to:
{\it (a)\/} LOG map, {\it (b)\/} TWB map. } \label{fig:planes}
\end{minipage}
\end{figure*}

As an application of the above quantifiers we study two usual
randomization procedures by means of the representation plane
depicted in Fig. \ref{fig:planes}, employing a quantifier
depending on $\mu(x)$ as $x$-axis ($H^{(hist)}$ is selected) and a
quantifier depending on $r_{mix}$ as $y$-axis ($DET$ is selected).
The first procedure is \textbf{Skipping} and the second one is
\textbf{Discretization} (see \cite{Demicco2008}).
\textbf{Skipping} has been used as a randomization procedure for
piecewise linear maps. The representation plane evidences the fact
that  this procedure is better than \textbf{Discretization}
because these maps already have the ideal invariant measure (they
have $H^{(hist)}=1$) and only $r_{mix}$ must be diminished to get
the ideal PRNG.   Fig. \ref{fig:planes} (b) shows that
\textbf{Discretization} is a better procedure for LOG because the
ideal point $[1,0]$ is not reached by \textbf{Skipping}.

\section{Conclusions}
\label{sec:SUMM}
In summary we were able to show here that:
\begin{enumerate}
    \item Two classes of quantifiers are required for the evaluation
         of the quality of a PRNG:
         {\it a)\/} \textit{quantifiers depending on $r_{mix}$ only (and not on $\mu{(x)}$}, like $H^{(hist)}$, $C^{(hist)}$, and $RR$ and
         {\it b)\/} \textit{quantifiers depending on $\mu{(x)}$ only (and not  on $r_{mix}$)}, as
         $C^{(BP)}$, $H^{(BP)}$, $DET$, $L$ and $ENTR$.
    \item Intrinsic computation quantifiers are dependent on both
    $\mu(x)$ and $r_{mix}$ and then they are not convenient for
    PRNG-analysis with our methodology.
    \item Representation planes with one quantifier of
    each class as coordinate axis allow for different chaotic PRNGs to be compared to each other so as to determine
    the best one. Furthermore, these representation planes permit one to judiciously select the best randomizing procedure.
\end{enumerate}

Our present results are consistent with those of previous works
\cite{Gonzalez2005,Larrondo2005,Demicco2008}.

\section{Acknowledgments.}
This work was partially  supported  by  the  Consejo  Nacional de
Investigaciones Cient\'{\i}ficas y T\'ecnicas (CONICET), Argentina
(PIP 5569/04, PIP 5687/05, PIP 6036/05), ANPCyT, Argentina (PICT
11-21409/04) and Universidad Nacional de Mar del Plata.
OAR gratefully acknowledge support from Australian Research Council
(ARC) Centre of Excellence in Bioinformatics, Australia.
%
%
%

\bibliographystyle{unsrt}
\bibliography{xbibWEB}

\end{document}